\documentclass[a4paper]{mem_sissa}
\usepackage{natbib}
\usepackage{graphicx}
\usepackage[a4paper]{hyperref}
\begin{document}
\title{NGC 6822: detection of variable stars with ISIS2.1
}
\author{L. Baldacci \inst{1,2}, G. Clementini \inst{2}, E.V. Held \inst{3}
     \and L. Rizzi\inst{3,4} 
}
\offprints{L. Baldacci}
\mail{via Ranzani 1, 40127 Bologna}
\institute{Universit\`a  di Bologna, via Ranzani 1, 40127 Bologna, Italy \\
 \and  INAF - Osservatorio Astronomico di Bologna, via Ranzani 1, 40127 
       Bologna, Italy \\  
 \and  INAF - Osservatorio Astronomico di Padova, vicolo dell'Osservatorio 5,
       35122 Padova, Italy \\
 \and  Universit\`a di Padova, vicolo dell'Osservatorio 2, 35122 Padova, 
       Italy}

\abstract{Preliminary results on the detection of a conspicuous 
number of variable stars in the dwarf 
irregular galaxy NGC 6822 are presented. We stress the need   
for packages specifically designed to the search of variables in distant
galaxies and/or when data are affected by a high level of crowding.

\keywords{Variable stars -- Local Group Galaxies -- NGC 6822}
   }
   \authorrunning{L. Baldacci et al.}
   \titlerunning{NGC 6822: detection of variable stars with ISIS2.1}
   \maketitle
%

\section{Introduction}
Variability involves many phases of stellar evolution: variable stars are 
both tracers of stellar populations and tests for stellar evolution 
theories. \\
As a part of the ongoing program on the study of variable stars in Local 
Group Galaxies carried out at the Bologna and Padua Observatories, we are 
studying the dwarf irregular galaxy NGC 6822. We obtained time series
observations (36 V and 11 B 15-min exposures for a total 
of 15 hours, spread on 3 nights) with the Very Large Telescope (VLT, 
Paranal-Chile) in August 2001. \\  
Traditionally the detection of variable stars is performed with the 
scatter diagram, which plots the mean magnitude of each star in the field 
and its standard deviation. 
Candidate variables are picked up from this diagram as stars whose 
standard deviation is larger than $3\sigma$, (where $\sigma$ is the rms of 
``bona-fide'' constant stars at the same magnitude level.)
Usually structures like fingers, corresponding to the region of classes of 
variables of constant mean magnitude, are seen in the scatter diagram, 
as the RR Lyrae finger and the Cepheid finger. 
This method works very well in globular clusters and nearby fields,
while does not work in distant galaxies or very crowded areas, like our 
NGC 6822 field. In very crowded conditions the scatter diagram is 
dominated by photometric errors, the typical structures are not visible 
and the efficiency of the detection rate decreases dramatically. 
\begin{figure*}
\centering
\includegraphics[width=\textwidth]{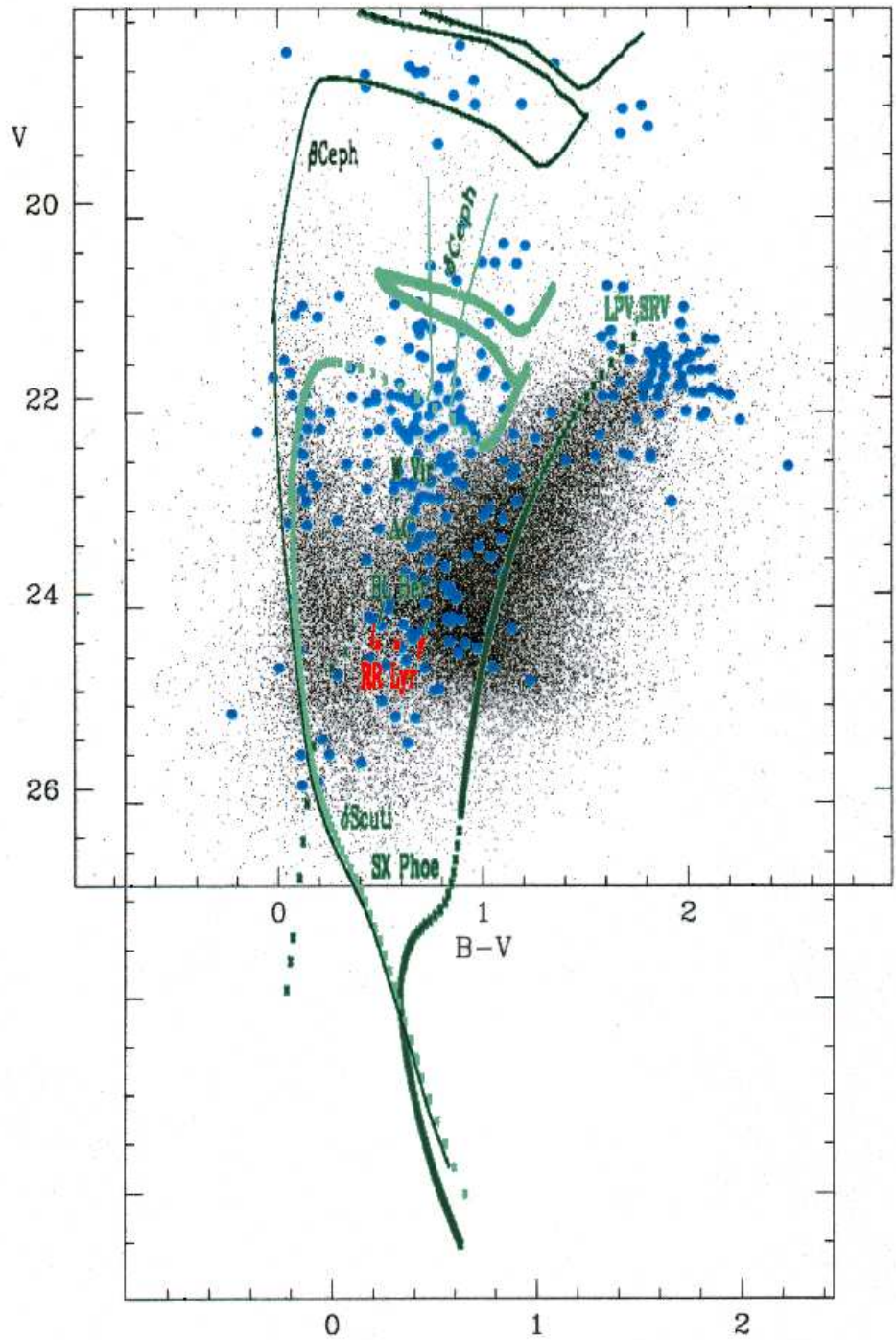}
\caption{Color magnitude diagram of NGC 6822. Big circles (blue in the electronic
version of the paper) mark candidate variable stars detected by ISIS2.1. 
Superimposed (in green) are the theoretical loci of the different types
of variable stars (from Marconi 2001). 
A reddening $\rm{E(B-V)}=0.25$ mag was assumed according to 
Schlegel et al. (1998).}
\label{fig:HR}
\end{figure*}

\section{ISIS2.1 and data reduction.}
ISIS2.1 \citep{alard} is a package specifically designed for the detection of 
variable stars in crowded fields. It is based on an optimal image subtraction 
technique: in the subtraction of two images of the 
same field all constant stars will cancel out and only variables will remain.
Two basic conditions must be satisfied to optimize the  
subtraction: the images must be geometrically and physically 
aligned. The former condition requires that all images have
the same coordinate system; the latter condition requires that 
all images have the same seeing condition and exposure time. \\
ISIS2.1 is designed to achieve all these requirements before going on with 
the subtraction of the images.
The procedure consists in the following steps: (1)geometric alignment of the 
image series and re-mapping of the images to the same grid; (2)construction 
of the reference image obtained by stacking a suitable subset of images, and 
convolution of the reference image with a kernel to adjust for seeing 
variations and geometrical distortions of
individual frames; (3)subtraction of each individual frame from the convolved 
reference image; (4)photometry of the variable objects on the difference 
images, in terms of differential fluxes.     
Light curves in a magnitude scale are then obtained through the photometric
reduction of the reference image. Reduction of the reference 
image was performed using DAOPHOT and ALLFRAME (Stetson 1994).\\ 
ISIS2.1 improves the variable stars search in two different ways, 
with respect to the traditional techniques: increases the efficiency of 
the detection rate, and improves the quality of the light curves. 
Each of these two points will be analysed in detail in the next paragraphs.  

\section{Efficiency of the detection rate.}
ISIS2.1 detected 450 candidate variables in NGC 6822. So far 
we have studied a subsample of about 100 of them (all candidate variables 
with v$\geq 13.5$ mag and ten of the brighter ones). About 90\% of them turned 
out to be real variable stars, they are mainly RR Lyrae stars and 
Anomalous Cepheids.
Less than 15\% of them would have been detected
with a traditional technique, since most of the real variables were found to
lie in regions of the scatter diagram dominated by photometric errors. \\ 
Figure \ref{fig:HR} shows the color magnitude diagram of NGC 6822, candidate 
variable objects are marked by big circles, solid lines are the theoretical 
loci occupied by different types of variables (from Marconi 2001). The figure 
was obtained superimposing the theoretical tracks to
NGC 6822 color magnitude diagram and shifting the x-axis to account for 
the redding effect ($\rm{E(B-V)}=0.25$ mag according to Schlegel et al. 1998).
All regions where theory predicts the presence of variables are 
well populated. The instability strip from Classical Cepheids  
to the RR Lyrae region can be clearly seen. 
Many variables are found also at the tip of the Red Giant Branch where 
are expected to lie long period variables ($\rm{P}\geq 50$ days), like 
Mira's and Semiregulars. Similar results were found by 
Bersier \& Wood (2002) in Fornax. They concluded that some of these stars 
are long period 
variables, while others are giants that vary with short periods and small 
amplitudes because of instabilities in their external layers. 
Some variables are also found on the main sequence of NGC 6822: 
they could be eclipsing binary systems or stars that pulsate on the main 
sequence, like $\beta$ Cephei stars. Study of the complete sample of candidate 
variables is in progress.
\begin{figure*}[t]
\centering
\includegraphics[bb=19 243 567 698, width=\textwidth]{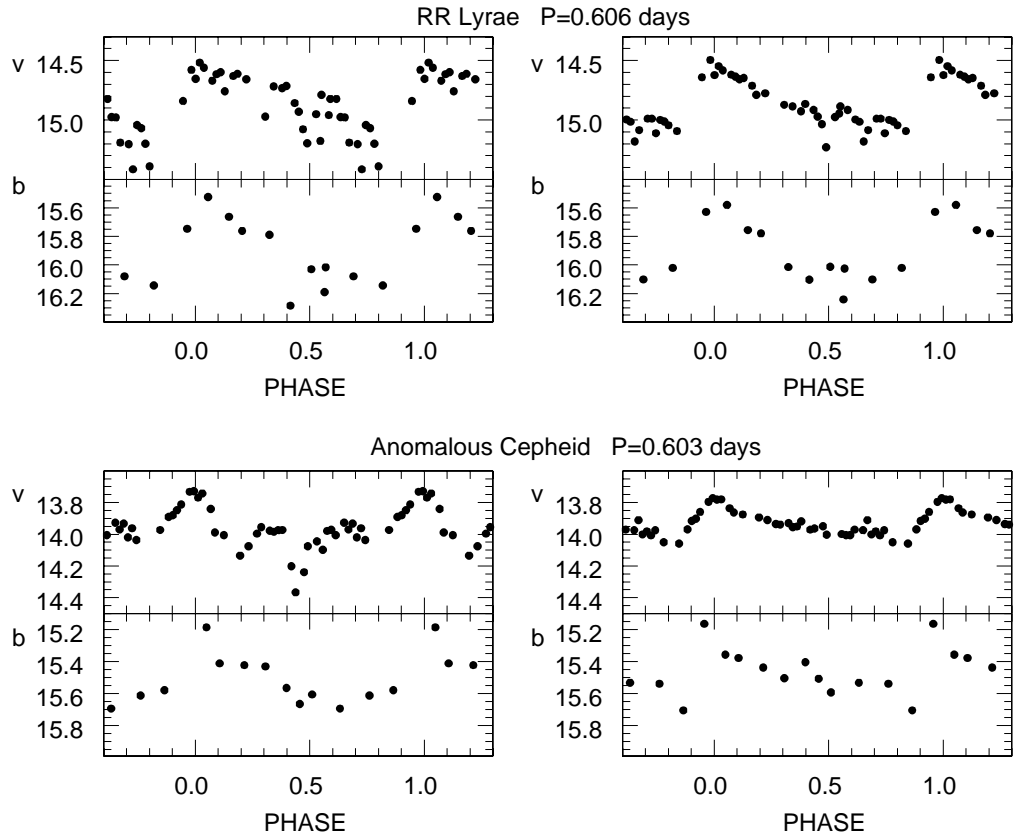}
\caption{Light curves of an Anomalous Cepheid and an RR Lyrae star of NGC 6822. 
Left panels: light curves obtained with DAOPHOT. 
Right panels: light curves obtained with ISIS2.1.}
\label{fig:curve2}
\end{figure*}

\section{Quality of light curves.}
Figure \ref{fig:curve2} shows the light curves for two of our variables. 
Light curves on the left panels were obtained with DAOPHOT, 
those on the right panels are from ISIS2.1.
The good quality of light curves allowed us the discovery of Anomalous 
Cepheids with very small amplitude ($0.1\div0.3$ mag). These variables 
would have never been recognized because of the very large scatter of their 
traditional DAOPHOT photometry.

\section{Conclusions.}
The really good performances of ISIS2.1 allowed us the detection for the 
first time of RR Lyrae stars and small amplitude Anomalous Cepheids in 
NGC 6822. RR Lyrae stars mark the Horizontal Branch level (otherwise not 
visible on the color magnitude diagram). The presence of the Horizontal 
Branch demonstrates without any doubt that NGC 6822 started forming 
stars on an early epoch: $\rm{t}\geq 10$ Gyr.

\bibliographystyle{aa}

\end{document}